\documentclass[12pt]{article}
\usepackage[latin1]{inputenc}

\usepackage{amsmath}
\usepackage{color}
\usepackage{amsfonts}
\usepackage{amssymb}
\usepackage[mathscr]{euscript}
\usepackage{graphicx}
\usepackage{geometry}
\usepackage{amssymb,epsfig,subfigure}
\usepackage{hyperref}
\usepackage{comment}
\usepackage{tabularx}
\usepackage{bm}
\usepackage{euscript}
\usepackage{graphicx}
\usepackage[dvipsnames]{xcolor}
\usepackage{amsfonts}
\usepackage{exscale}
\usepackage{amsbsy}
\usepackage{subfigure}
\usepackage{textcomp}
\usepackage{comment}
\usepackage{hyperref}
\usepackage{slashed}
\usepackage{authblk}
\usepackage{tabularx}
\usepackage{euscript}
\usepackage{graphicx}
\usepackage{color}
\usepackage{amsfonts}
\usepackage{exscale}
\usepackage{amsbsy}
\usepackage{subfigure}
\usepackage{textcomp}
\usepackage{comment}
\usepackage{hyperref}
\usepackage{bm}
\usepackage{wrapfig}
\usepackage[font=footnotesize,labelfont=bf]{caption}
\usepackage{ulem} 

\def\ba#1\ea{\begin{align}#1\end{align}}
\def\bg#1\eg{\begin{gather}#1\end{gather}}
\def\bm#1\em{\begin{multline}#1\end{multline}}
\def\bmd#1\emd{\begin{multlined}#1\end{multlined}}

\newcommand{\be}{\begin{equation}}
\newcommand{\ee}{\end{equation}}
\newcommand{\bea}{\begin{eqnarray}}
\newcommand{\eea}{\end{eqnarray}}

\newcommand{\pd}{\partial}
\newcommand{\bs}{\boldsymbol}
\newcommand{\matleft}{\left(\begin{array}}
\newcommand{\matright}{\end{array}\right)}
\newcommand{\Tr}{\operatorname{Tr}}

\usepackage[numbers,sort&compress]{natbib}
\def\simge{
    \mathrel{\rlap{\raise 0.511ex 
        \hbox{$>$}}{\lower 0.511ex \hbox{$\sim$}}}}

\def\simle{
    \mathrel{\rlap{\raise 0.511ex 
        \hbox{$<$}}{\lower 0.511ex \hbox{$\sim$}}}}

\makeatletter
\renewcommand\section{\@startsection {section}{1}{\z@}%
                                 {-3.5ex \@plus -1ex \@minus -.2ex}
                                   {2.3ex \@plus.2ex}%
                                   {\normalfont\large\bfseries}}
\renewcommand\subsection{\@startsection{subsection}{2}{\z@}%
                                   {-3.25ex\@plus -1ex \@minus -.2ex}%
                                     {1.5ex \@plus .2ex}%
                                     {\normalfont\bfseries}}
\renewcommand\subsubsection{\@startsection{subsubsection}{3}{\z@}%
                                   {-3.25ex\@plus -1ex \@minus -.2ex}%
                                     {1.5ex \@plus .2ex}%
                                     {\normalfont\itshape}}
\makeatother

\def\pplogo{\vbox{\kern-\headheight\kern -29pt
\halign{##&##\hfil\cr&{\ppnumber}\cr\rule{0pt}{2.5ex}&\ppdate\cr}}}
\makeatletter
\def\ps@firstpage{\ps@empty \def\@oddhead{\hss\pplogo}%
  \let\@evenhead\@oddhead 
}
\def\maketitle{\par
 \begingroup
 \def\thefootnote{\fnsymbol{footnote}}
 \def\@makefnmark{\hbox{$^{\@thefnmark}$\hss}}
 \if@twocolumn
 \twocolumn[\@maketitle]
 \else \newpage
 \global\@topnum\z@ \@maketitle \fi\thispagestyle{firstpage}\@thanks
 \endgroup
 \setcounter{footnote}{0}
 \let\maketitle\relax
 \let\@maketitle\relax
 \gdef\@thanks{}\gdef\@author{}\gdef\@title{}\let\thanks\relax}
\makeatother

\hypersetup{
    bookmarks=true,         
    unicode=false,          
    pdftoolbar=true,        
    pdfmenubar=true,        
    pdffitwindow=false,     
    pdfstartview={FitH},    
    pdftitle={LG Theories of Non-Abelian QH States},    
    pdfauthor={},     
    pdfsubject={Subject},   
    pdfcreator={},   
    pdfproducer={}, 
    pdfkeywords={keyword1} {key2} {key3}, 
    pdfnewwindow=true,      
    colorlinks=true,       
    linkcolor=OliveGreen, 
    citecolor=NavyBlue,        
    filecolor=magneta,      
    urlcolor=cyan           
}

\numberwithin{equation}{section}

\newcommand*\samethanks[1][\value{footnote}]{\footnotemark}

\newcommand\blfootnote[1]{%
  \begingroup
  \renewcommand\thefootnote{}\footnote{#1}%
  \addtocounter{footnote}{-1}%
  \endgroup
}

\begin{document}

\setcounter{page}0
\def\ppnumber{\vbox{\baselineskip14pt
}}

\def\ppdate{
} \date{\today}

\title{\Large \bf Landau-Ginzburg Theories of Non-Abelian Quantum Hall States from Non-Abelian Bosonization}
\author{Hart Goldman$^\psi$, Ramanjit Sohal$^\phi$, and Eduardo Fradkin}
\affil{ \it Department of Physics and Institute for Condensed Matter Theory,\\  \it University of Illinois at Urbana-Champaign, \\  \it 1110 West Green Street, Urbana, Illinois 61801-3080, USA}
\maketitle
\begin{abstract}
It is an important open problem to understand the landscape of non-Abelian fractional quantum Hall phases which can be obtained starting from physically motivated theories of Abelian composite particles. We show that progress on this problem can be made using recently proposed non-Abelian bosonization dualities in 2+1 dimensions, which morally relate $U(N)_k$ and $SU(k)_{-N}$ Chern-Simons-matter theories. The advantage of these dualities is that regions of the phase diagram which may be obscure on one side of the duality can be accessed by condensing local operators on the other side. 
Starting from parent Abelian states, we use this approach to construct Landau-Ginzburg theories of non-Abelian states through a pairing mechanism. In particular, we obtain the bosonic Read-Rezayi sequence at fillings $\nu=k/(kM+2)$ by starting from $k$ layers of bosons at $\nu=1/2$ with $M$ Abelian fluxes attached. The Read-Rezayi states arise when $k$-clusters of the dual non-Abelian bosons condense. We extend this construction by showing that $N_f$-component generalizations of the  Halperin $(2,2,1)$ bosonic states have dual descriptions in terms of $SU(N_f+1)_1$ Chern-Simons-matter theories, revealing an emergent global symmetry in the process. Clustering $k$ layers of these theories yields a non-Abelian $SU(N_f)$-singlet state at filling $\nu = kN_f / (N_f + 1 + kMN_f)$.\blfootnote{$^{\psi\,\leftrightarrow\,\phi}$ These authors contributed equally to the development of this work.} 
\end{abstract}
\bigskip
\newpage

{
\hypersetup{linkcolor=black}
\tableofcontents
}
\thispagestyle{empty}

\newpage
\setcounter{page}{1}

\section{Introduction}
Two-dimensional charged quantum fluids in a strong magnetic field exhibit an impressive array of topologically ordered incompressible states at partial Landau level (LL) fillings $\nu$, in what is known as the fractional quantum Hall (FQH) effect. Of these states,  
those exhibiting Abelian topological order are readily understood through the notion of flux attachment \cite{Wilczek-1982}, which exactly relates fermions or (hard-core) bosons at fractional LL filling to a theory of either composite fermions \cite{Jain-1989,Lopez-1991} or bosons \cite{Zhang-1989} in a reduced magnetic field. After flux attachment, the Abelian FQH states may either be viewed as integer quantum Hall (IQH) states of composite fermions or as a condensate of composite bosons governed by a Landau-Ginzburg (LG) theory. 

Despite the success over the past several decades in understanding the Abelian FQH states, an understanding of the dynamics which can lead to {\it non-Abelian} FQH states has remained elusive. Such states cannot arise directly from the application of flux attachment, which is by definition Abelian. For example, while it is believed that the observed $\nu=5/2$ FQH plateau is a non-Abelian state arising from composite fermion pairing \cite{Read2000}, the origin and nature of the pairing instability leading to this state continues to be debated, with seemingly contradictory results between experiment and numerics \cite{Banerjee2018,Storni2010,Zaletel2015,Rezayi2017,Mishmash2018}. 
Nevertheless, assuming a particular pairing channel, 
a non-Abelian phase appears quite naturally \cite{Moore1991,Read2000}. 



Unfortunately, this physical picture does not appear to translate simply to the other proposed non-Abelian states, such as the Read-Rezayi (RR) states  \cite{Read1999}. Wave functions for these states can be constructed using conformal field theory (CFT) techniques \cite{Moore1991}, but it is not clear which of these states can be obtained 
starting from a (physically motivated) field theory of composite particles. To make matters worse, the wave functions for generic non-Abelian states are typically characterized by clustering of more than two particles \cite{Read1999,Ardonne-2004}. Na\"{i}vely, from perturbative scaling arguments, such states could not arise unless the clusters with fewer particles are disallowed by symmetry. Most theories of interest do not appear to have such a symmetry, implying that non-perturbatively strong interaction effects are required to give rise to such states. While we note that projective/parton constructions can be used to formulate effective bulk theories of non-Abelian states \cite{Wen1991,Wen1999,Barkeshli2010}, in such constructions the electron operator is fractionalized by hand, and it must be taken by fiat that the fractionalized degrees of freedom are deconfined. 
Consequently, although the projective approach can formally generate many candidate states, it does not shed much light on their dynamical origin.

Recent progress in the study of non-Abelian Chern-Simons-matter theories in their large-$N$ (``planar'') limit \cite{Aharony2012,Giombi2012} has led to the proposal of non-Abelian Chern-Simons-matter theory dualities by Aharony \cite{Aharony2016}, which take the shape of level-rank dualities. Along with  the Abelian web of dualities they imply \cite{Seiberg2016,Karch2016}, these dualities constitute tools with which it may be possible to make non-perturbative progress on the above problem. Such dualities can relate theories of Abelian composite particles to theories of non-Abelian monopoles, and they have led to progress on several important problems in condensed matter physics \cite{Son2015,Wang2015,Metlitski2016,Radicevic2016,Goldman2017,Hui2019,Wang2017a,Thomson2017,Goldman2018a}. Of particular importance for us, pairing deformations of a dual non-Abelian theory can lead to non-Abelian topological phases which appear inaccessible to the original Abelian theory, in which this pairing corresponds to a highly non-local product of monopole operators. 

Our strategy is to use these non-Abelian dualities to begin to map the landscape of non-Abelian topological phases accessible from a ``composite particle'' picture, by way of ``projecting down'' from a multi-layer parent Abelian state. This type of approach, in which the transition to the non-Abelian phase can be physically interpreted as being driven by interlayer tunneling \cite{Wen2000,Read2000,Cabra2000,Cabra2001,Rezayi2010,Barkeshli2010a,Barkeshli2011,Vaezi2014} or pairing \cite{Fradkin1999,Ardonne1999}
, has formed the foundation of several lines of attack on the non-Abelian FQH problem. Such projections have been implemented at the formal level of the edge CFT (``ideal") wave function \cite{Cappelli2001,Ardonne2003} and in coupled wire constructions \cite{Teo2014,Fuji2017}. 
Numerical studies of bilayer systems have also lent support to this idea \cite{Papic2010,Peterson2015,Geraedts2015a,Liu2015,Zhu2015,Zhu2016,Crepel-2019}.  
However, 
a robust bulk LG description of generic non-Abelian FQH states continues to be lacking. In one major attempt to fill this gap, the authors of Ref. \cite{Fradkin1999} constructed a non-Abelian LG theory for a subset of the bosonic RR states by considering layers of $\nu=\frac{1}{2}$ (bosonic) Laughlin states. Using the well-known level-rank duality of the (gapped) bulk Chern-Simons topological quantum field theory (TQFT) \cite{Naculich-1990,Naculich-1990b,Camperi-1990} (see Ref. \cite{Fradkin-2013} for a review), the authors motivated a description of these states involving $SU(2)$ Chern-Simons gauge fields coupled to scalar matter in the adjoint (matrix) representation, obtaining the non-Abelian QH state by pairing across the different layers. In this approach, the anyon content of the non-Abelian state is furnished by the vortices of the pairing order parameter. While this construction is conceptually appealing, it does not originate from a duality satisfied by the parent Abelian LG theory, which describes a quantum critical point, but, rather, a duality satisfied only deep in the gapped Abelian FQH phase. Moreover, in order to give the anyons electric charge in this approach, it is necessary for the external electromagnetic field to couple to the $U(1)$ subgroup of the full non-Abelian gauge group, explicitly breaking the larger gauge invariance.

\begin{figure}
  \centering
    \includegraphics[width=0.85\textwidth]{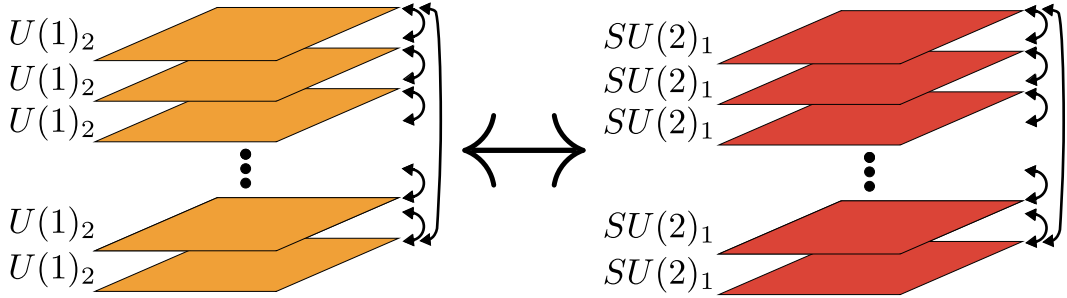}
    \caption{A schematic of our construction of LG theories for the RR states. 
    $k$ copies of the $\nu=\frac{1}{2}$ Laughlin state coupled to scalars (left) 
    are dual to $k$ copies of $SU(2)_1$ coupled to scalars (right)
    . The $SU(2)_k$ Read-Rezayi states are obtained in the dual, non-Abelian language via pairing of the layers, represented by double-headed arrows. In the original Abelian theory, these correspond to non-local, monopole interactions. 
    } \label{fig:cartoon}
\end{figure}

Using the non-Abelian bosonization dualities, we construct LG theories of the full bosonic RR sequence at filling fractions $\nu=k/(kM+2)$, $k,M\in\mathbb{Z}$, which do not suffer from these problems. These theories are obtained by starting with $k$ layers of $\nu=1/2$ bosonic QH states, using the dualities to obtain a LG theory of non-Abelian composite bosons, and attaching $M$ fluxes to the resulting theory. 
For example, we obtain a LG theory of the bosonic $\nu=1$ Moore-Read state consisting of two layers of bosons $\phi_n$, $n=1,2$, which we call the ``composite vortices,'' each at their Wilson-Fisher fixed point and coupled in the {\it{fundamental}} representation to a $SU(2)$ gauge field $a_n$,
\be
\label{eq: MR Theory B intro}
\mathcal{L}=\sum_{n=1}^2\left[|D_{a_n-A\bs{1}/2}\,\phi_n|^2-|\phi_n|^4+\frac{1}{4\pi}\Tr\left(a_nda_n-\frac{2i}{3}a_n^3\right)\right]-\frac{1}{4\pi}AdA\,.
\ee
where $D_{a_n-A\bs{1}/2}=\pd-i(a_n^b t^b-A\bs{1}/2)$ is the covariant derivative, we use the notation $AdB=\varepsilon^{\mu\nu\lambda}A_\mu\pd_\nu B_\lambda$, $t^b=\sigma^b/2$ are the $SU(2)$ generators, and $\bs{1}$ is the $2\times 2$ identity matrix. We use the notation $-|\phi|^4$ to denote tuning to the Wilson-Fisher fixed point. Although the gauge fields $a_n$ are non-Abelian, the topological phase accessed by simply gapping out the composite vortices will only support excitations with \textit{Abelian} statistics. For a $SU(N)$ gauge group, non-Abelian statistics require the presence of a Chern-Simons term at level greater than one. To obtain the non-Abelian FQH state, we condense clusters of the non-Abelian composite vortices across the layers (see Fig. \ref{fig:cartoon}), in this case condensing $\phi_1^\dagger\phi_2$ without condensing $\phi_{1},\phi_2$ individually. This Higgses the linear combination $a_1-a_2$ of the $SU(2)_1$ gauge fields, causing the bilayer $SU(2)\times SU(2)$ gauge group to be broken down to its diagonal $SU(2)$ subgroup. The Chern-Simons levels of the resulting gapped phase add, leading to the desired $SU(2)_2$ Chern-Simons theory at low energies (the subscript refers to the Chern-Simons level). We will show below that 
the composite vortices individually have the proper quantum numbers to fill out the anyon spectrum of the theory. The clarity of the topological content of the non-Abelian states is a general advantage of the bosonic LG approach. However, alternative descriptions of non-Abelian FQH states involving dual non-Abelian composite \textit{fermions} are also possible. We plan to describe this complementary perspective in future work.

In addition to the the RR states, by considering $N_f$-component generalizations of the Halperin (2,2,1) spin-singlet states on each layer, we are able to generalize this approach to construct bulk LG descriptions of generalized  non-Abelian $SU(N_f)$-singlet (NASS) states at fillings \cite{Reijnders2002,Fuji2017},
\begin{align}
\nu = \frac{kN_f}{N_f+1+kMN_f}\,, \qquad k, N_f, M \in \mathbb{Z}\, ,\label{eqn:generalized NASS fillings}
\end{align}
which are bosonic (fermionic) for $M$ even (odd). These states generalize the clustering properties of the RR states to $N_f$-component systems and, as their name suggests, are singlets under $SU(N_f)$ rotations. Indeed, for $N_f=1$, these states reduce to the RR states while for $N_f=2$, they describe the non-Abelian spin singlet (also NASS) states of Ardonne and Schoutens \cite{Ardonne1999,Ardonne2001}. These generalized NASS states morally possess $SU(N_f+1)_k$ topological order, 
and so support anyons obeying the fusion rules of Gepner parafermions \cite{Gepner1987}, generalizations of the $\mathbb{Z}_k$ parafermions \cite{Fradkin1980} found in the RR states.
Although the physical relevance of an $N_f$-component FQH state may seem dubious for larger values of $N_f$, the generalized NASS states provide candidate ground states in systems of cold atoms \cite{Reijnders2002,Reijnders2004} and fractional Chern insulators \cite{Sterdyniak2013}.
In building LG theories of these states, we find a new duality relating (A) $N_f$ Wilson-Fisher bosons coupled to $U(1)$ Chern-Simons gauge fields with Lagrangian given by the $N_f$-component generalization of the Halperin $(2,2,1)$ $K$-matrix theory to (B) a $SU(N_f+1)_1$ Chern-Simons theory coupled to $N_f$ Wilson-Fisher bosons in the fundamental representation. This non-Abelian dual description makes manifest the emergent $SU(N_f)$ global symmetry and reflects the fact that the edge theory of the $N_f$-component $(2,2,1)$ state supports an $SU(N_f+1)_1$ Kac-Moody algebra.

The remainder of this work is organized as follows. We begin in Section \ref{sec: review} by elaborating on the motivation for our construction both from the perspective of wave functions and that of the earlier Landau-Ginzburg approach of Ref. \cite{Fradkin1999}. We then proceed to our analysis in Section \ref{section: Read Rezayi Bosonization} of the RR states using non-Abelian bosonization, resolving the lingering issues of the LG construction of Ref. \cite{Fradkin1999}. We then extend our construction to the generalized NASS states in Section \ref{sec: generalized NASS}. Future directions are discussed in Section \ref{sec: discussion}. 

\section{``Projecting Down'' to Non-Abelian States}
\label{sec: review}


\subsection{Perspective from the Boundary: Wave Functions and their Symmetries}
\label{sec: RR wave fncs}
If we wish to construct a LG description of non-Abelian FQH states involving pairing between Abelian states, it is first necessary to identify which Abelian states to pair. 
Such states can be motivated 
by considering ``ideal" wave functions. These can be constructed from certain correlation functions, known as conformal blocks, of the edge CFT. In this language, 
the strategy of obtaining non-Abelian states from parent Abelian states through ``projecting down" is well established \cite{Cappelli2001}. 

Consider for example the bosonic RR states at $\nu=k/2$. The ideal wave functions of these states are defined as the ground states of ideal $k+1$-body Hamiltonians, which can be shown to be given by the conformal blocks of the $SU(2)_k$ Wess-Zumino-Witten (WZW) CFT \cite{Read1999}. This tells  
us that the RR wave functions describe FQH states with edges governed by $SU(2)_k$ WZW theories \cite{Moore1991}, corresponding in the bulk to 
a $SU(2)_k$ Chern-Simons gauge theory \cite{Witten1989}. 
A natural way to obtain the ideal wave functions for the $\nu=k/2$ RR states uses the state with $k=1$ -- the $\nu=1/2$ bosonic Laughlin state, which is Abelian -- as a building block \cite{Cappelli2001}. This state is described by the wave function 
\begin{align}
\Psi_{1/2}(\{z_{i}\}) = \prod_{i<j} (z_{i} - z_{j} )^{2}  e^{-\frac{1}{4}\sum_{i} |z_i|^2}, \label{eqn:laughlin-wavefn}
\end{align}
where $z_j = x_j + iy_j$ denotes the complex coordinates of the $j^{\mathrm{th}}$ particle (a boson). 
The $\nu=k/2$ RR wave functions may be obtained from this one by ``clustering" bosons across $k$ copies of this state. This corresponds to taking $N=k m$ bosons, dividing them into $k$ groups, writing down a $\nu=\frac{1}{2}$ Laughlin wave function for each group, multiplying them together, and then symmetrizing over all possible assignments of 
bosons to groups. The resulting wave function is represented as  
\begin{align}
\Psi_{k}(\{z_i\}) = \mathcal{S}_k \left[ \prod_{i=0}^{k-1} \Psi_{1/2}(z_{1+iN/k},\dots,z_{(i+1)N/k})\right], \label{eqn:RR-wavefn}
\end{align}
where $\mathcal{S}_k$ denotes symmetrization. It can be shown that this wave function is equivalent to that first proposed by Read and Rezayi \cite{Read1999} and exhibits the correct clustering properties: the wave function does not vanish unless the coordinates of $k+1$ bosons coincide. The RR wave functions for general $k$ and $M$ are obtained by multiplying Eq. \eqref{eqn:RR-wavefn} by a $\nu=\frac{1}{M}$ Laughlin factor.

The relation between the $k=1$ and the $k>1$ RR wave functions suggests that 
it should be possible to construct such a LG theory by considering $k$ copies of the effective theory of the ({\it Abelian}) $k=1$ state, the first attempt at which we describe in the next subsection. 
That a state with $SU(2)_k$ topological order can be obtained from the Abelian $\nu=\frac{1}{2}$ Laughlin state is also made plausible by the fact that the latter has an alternative description as an $SU(2)_1$ Chern-Simons theory. This is a consequence of the level-rank duality between $U(1)_2$ and $SU(2)_1$, which is reflected in the above description by the fact that the $\nu=\frac{1}{2}$ wave function can be obtained from the $SU(2)_1$ WZW CFT 
\cite{Fradkin1999,Read1990,Frohlich1991,Fuji2017}. We review this level-rank duality in the subsection below.

\subsection{Perspective from the Bulk: Early LG Theories from Level-Rank Duality} \label{sec: Early LG}

To approach the problem of constructing a bulk description of the Read-Rezayi states, the authors of Ref. \cite{Fradkin1999} 
sought to obtain a non-Abelian Landau-Ginzburg theory of the $\nu=k/2$ RR states by also considering $k$ layers of $\nu=1/2$ bosonic Laughlin states, or $U(1)_{2}$ Chern-Simons theories and 
recognizing that each $U(1)_2$ theory is level-rank dual to a $SU(2)_{1}$ theory. 
They therefore conjectured that an alternate LG description was possible, one involving scalar matter coupled to $SU(2)_{1}$ gauge fields. These scalars could then pair and lead to the symmetry breaking pattern,
\be
\label{eq: symmetry breaking pattern}
SU(2)_{1}\times\cdots\times SU(2)_{1}\rightarrow SU(2)_{k}\,.
\ee 
What remained was to (1) determine how the scalars transformed under $SU(2)$ and how they coupled to the physical background electromagnetic (EM) field, and (2) determine precisely how to pair these fields to obtain non-Abelian states.
 
For simplicity, we consider first the case of $k=2$, a bilayer of $\nu=1/2$ bosonic FQH liquids. This will constitute a parent state for the $\nu=1$ bosonic Moore-Read state.
To motivate the level-rank duality to a non-Abelian representation, we again consider the edge physics. The edge theory of the $U(1)_{2}$ state is one of a chiral boson, 
\be
\mathcal{L}_{\mathrm{edge}}=\frac{1}{4\pi\nu}\,\pd_x\varphi\,(\pd_t\varphi-v\pd_x\varphi)\, ,
\ee
where $\varphi$ has compactification radius $R=1$ and $\nu=1/2$. 
 The charge density is therefore $\rho=\frac{1}{2\pi}\pd_x\varphi$. The local particles (i.e. the physical bosons) of this theory are represented by the  vertex operators,
\be
\psi_1=e^{i\varphi / \nu}.
\ee
In addition, the theory hosts anyonic quasiparticles, which are semions of charge $1/2$ and correspond to the vertex operators 
\be
\psi_{1/2}=e^{i\varphi}\,.
\ee
The $\psi_{1},\psi_1^\dagger,$ and $\rho$ operators all have the same scaling dimension and furnish a $SU(2)_1$ Kac-Moody algebra. This is a manifestation of the level-rank duality at the level of the edge CFT, and we can write the bulk theory on each layer as a $SU(2)_{1}$ gauge theory with gauge field $a_\mu=a_\mu^bt^b$, where $t^b$ are the generators of $SU(2)$. Importantly, the $\rho$ operator appears as the diagonal generator of $SU(2)$. Therefore, the authors guessed that in the LG theory the background EM field couples through a BF term to the Cartan component of the bulk $SU(2)$ gauge field\footnote{Note that, depending on context, we use $a^3$ to denote both the diagonal element of $a$ as well as $a\wedge a\wedge a$.},
\be
\mathcal{L}_{\mathrm{EM}}[a^a,A]=\frac{1}{2\pi}\varepsilon^{\mu\nu\lambda}A_\mu\pd_\nu a^3_\lambda\,.
\ee
This explicitly breaks gauge invariance and would indicate that the physical EM current is not conserved. We will eventually see in Section \ref{section: Read Rezayi Bosonization} that the new dualities will allow us to avoid this difficulty by granting us a gauge invariant way of coupling to the background electromagnetic field.

From this discussion, a natural guess for the matter variables for the bulk LG theory is a $SU(2)$ triplet on each layer consisting of boson creation and annihilation operators $B_n,B_n^\dagger$ and a boson number operator $B_n^3$ which essentially corresponds to the EM charge. Here $n=1,2$ is a layer index. If we write $B_n=B_n^1+iB_n^2$ with $B_n^{1,2}$ real, the adjoint field $B^a_n$ transforms like a vector under $SO(3)$. It is important to note, however, that any non-Abelian LG theory should be thought of as describing a (UV) quantum critical point proximate to the (IR) FQH state which shares universal features with the Abelian theory we started with. Since the level-rank duality is invoked deep in the FQH phase, it is a guess that these variables are the proper degrees of freedom at the UV quantum critical point (they may be alternatively understood as bound states -- we will see later on that this interpretation is more accurate). 
Nevertheless, pairing these fields will lead to both the desired symmetry breaking pattern \eqref{eq: symmetry breaking pattern} as well as the existence of solitons with non-Abelian statistics. 

The LG theory for the pairing of these fields can be explicitly constructed as follows. Each layer consists of a $B_n^a$ field minimally coupled to its own $SU(2)_1$ gauge field, 
\be
\label{eq: adjoint LG action}
\mathcal{L}_0[B_n,a_n]=\sum_{n=1,2}\left(|D_{a_n}B_n|^2+\frac{1}{4\pi}\Tr\left[a_nda_n-\frac{2i}{3}a^3_n\right]\right)+\cdots\,,
\ee
where we have suppressed Lorentz and $SU(2)$ indices, used the notation $AdC=\varepsilon^{\mu\nu\lambda}A_\mu\pd_\nu C_\lambda$, and defined the covariant derivative $D_{a_n}B_n\equiv\pd B^a_n-i\varepsilon^{abc}a_n^bB^c_n$.
The ellipsis refers to additional contact terms, Maxwell terms, etc. These are set up so that, taken individually, when each layer is at filling $\nu=1/2$, the diagonal color flux $b^3_I=\langle f^3_{I,xy}\rangle/2\pi$, vanishes. 

Although the $B_n$ fields are bosons, we assume that they do not condense. Rather, we consider pairing them using a method analogous to that of Jackiw and Rossi \cite{Jackiw1981}, who considered pairing Dirac fermions by coupling them to a scalar order parameter which mediates the pairing interaction. 
Let us introduce a field $\mathcal{O}^{ab}$  which transforms as an adjoint under each layer's $SU(2)$,
$\mathcal{O}\mapsto G_1^{-1}\,\mathcal{O}\,G_2$, where $G_{1},G_2\in SO(3)$. Here
we have used the fact that, as an adjoint field, $\mathcal{O}$ is blind to the $\mathbb{Z}_2$ centers of the two $SU(2)$ factors, and so effectively transforms under $SU(2)/\mathbb{Z}_2 \cong SO(3)$. The field $\mathcal{O}$ mediates a pairing interaction between the $B^a_I$ fields as follows,
\be
\label{eq: quartetting term}
\mathcal{L}_{\mathrm{pair}}=\lambda\,B^a_1\mathcal{O}^{ab}B^b_2\,.
\ee
We now require that $\mathcal{O}$ acquires a vacuum expectation value (VEV), which breaks $SU(2)\times SU(2)$ down to its diagonal subgroup $SU(2)_{\mathrm{diag}}$, implementing the constraint $a_1=a_2$. Any VEV equivalent to $\langle\mathcal{O}\rangle\propto \delta^{ab}$ is sufficient to achieve this. Therefore, in the final IR theory, the CS terms for $a_1$ and $a_2$ add, yielding a $SU(2)_2$ CS term, which describes precisely the $\nu=1$ bosonic Moore-Read state. The authors of Ref. \cite{Fradkin1999} then argued that, since the order parameter is valued on $[SO(3)\times SO(3)]/SO(3)$, that it can host non-trivial vortices which furnish the anyon content. This is in contrast to if we had chosen to pair fields in the fundamental representation, for which the order parameter has no non-trivial vortices. Finally, we note that because $\mathcal{O}$ is blind to the centers of the two original $SU(2)$ factors, the final gauge group is in fact $SU(2)_{\mathrm{diag}} \times \mathbb{Z}_2$. This means that the resulting topological order is not quite that of the $\nu=1$ bosonic Moore-Read state. We will elaborate on this point as well as the interpretation of the vortices in Section \ref{sec:Read Rezayi Quartetting}.

In spite of its successes, the LG theory described here has several problems. As mentioned above, the BF coupling between $a_n^3$ and the EM field $A$ explicitly breaks the $SU(2)$ gauge symmetry. In addition, the theory of adjoint fields \eqref{eq: adjoint LG action} cannot be the same as the Abelian LG theory of the original layers -- the theories have different phase diagrams and so do not represent the same fixed point. Moreover, the final gauge group after pairing is not just $SU(2)$ but includes additional discrete gauge group factors. 
Finally, it is not entirely obvious how to generalize this approach to the rest of the Read-Rezayi states and beyond. In this work, using non-Abelian boson-fermion dualities, we repair all of these problems.

\section{LG Theories of the RR States from Non-Abelian Bosonization \label{section: Read Rezayi Bosonization}}

\subsection{Setup}

Our setup for obtaining LG theories of the RR states is depicted in Figure \ref{fig:cartoon}. We again consider $k$ layers of bosonic quantum Hall fluids at $\nu=1/2$. 
The standard LG theory \cite{Zhang-1989} of these states consists of Wilson-Fisher bosons -- the Laughlin quasiparticles -- on each layer, denoted $\Phi_n$, with $n=1,\cdots,k$ being the layer index. Each of these fields is coupled to an {\it Abelian} $U(1)_2$ Chern-Simons gauge field $a_{n}$ as follows (the total gauge group is $[U(1)]^k$), 
\be
\mathcal{L}_A=\sum_{n}\left(|D_{a_{n}}\Phi_{n}|^2-|\Phi_{n}|^4+\frac{2}{4\pi}a_nda_{n}+\frac{1}{2\pi} A da_{n}\right)\,. \label{eqn: Abelian Theory A}
\ee
where again $-|\Phi|^4$ denotes tuning to the Wilson-Fisher fixed point and $D_{a_n}=\pd-ia_n$ is the covariant derivative.  
Since we wish to impose particle-hole symmetry on the bosons in the FQH state, these theories are relativistic. 
We take the background EM field $A_\mu$ to couple to the sum of the global $U(1)$ currents on each layer $j_{\mathrm{top}}=\frac{1}{2\pi}\sum_{n} da_{n}$, although we could have in principle coupled background fields to each of these currents individually \cite{Wen1995}. Notice that there is no continuous flavor symmetry manifest in $\mathcal{L}_A$ since each $\Phi_{n}$ couples to its own gauge field $a_{n}$. Being a theory of Laughlin quasiparticles, the Abelian quantum Hall state arises when the $\Phi$ fields are gapped, or $\rho_\Phi=\sum_{I,n}\langle i(\Phi^\dagger_{n}\overleftrightarrow{D}_{a_{n},t}\Phi_{n})\rangle=0$. We note here that throughout this paper we define the filling fraction with a minus sign $\nu=-2\pi\rho_e/B$, where $\rho_e$ is the physical EM chage and $B$ is the background magnetic field. 


We call the Abelian theory whose Lagrangian $\mathcal{L}_A$ is shown in Eq. \eqref{eqn: Abelian Theory A}, \textbf{Theory A}. In order to obtain a non-Abelian $SU(2)_k$ theory, our strategy is to invoke a non-Abelian duality to trade $\mathcal{L}_A$ for a theory of $k$ bosons which are charged under emergent non-Abelian gauge fields. Since these particles are non-Abelian analogues of the Laughlin quasiparticles (they are {\it gapped} in the Abelian QH state), we will refer to them as non-Abelian composite {\it vortices}. Indeed, we will see that these theories are the $k$-component generalizations of the theory of Eq. \eqref{eq: MR Theory B intro}. We call this non-Abelian theory \textbf{Theory B}. By pairing these fields across the layers, we will obtain the final $SU(2)_k$ theory. Thus, the non-Abelian FQH states we obtain can be interpreted as clustered states of the dual non-Abelian composite vortices, in analogy to the clustering interpretation of the wave functions. Moreover, from products of the non-Abelian vortex fields, analogues of the adjoint $B_n$ operators of Section \ref{sec: review} can be constructed and paired, leading to a ``quartetted" non-Abelian state. We now turn to a procedure for obtaining these dualities.

\subsection{A Non-Abelian Duality: $U(1)_2$ + bosons $\longleftrightarrow$ $SU(2)_1$ + bosons}
\label{subsec:numerology}
The non-Abelian dualities presented by Aharony \cite{Aharony2016} relate Chern-Simons theories coupled to complex scalar fields at their Wilson-Fisher fixed point to dual Chern-Simons theories coupled to Dirac fermions,
\bea
\label{eq: U/SU}
\text{$N_f$ scalars + $U(N)_{k,k}$} &\longleftrightarrow& \text{$N_f$ fermions + $SU(k)_{-N+N_f/2}$}\,, \\
\label{eq: SU/U}
\text{$N_f$ scalars + $SU(N)_k$} &\longleftrightarrow& \text{$N_f$ fermions + $U(k)_{-N+N_f/2,-N+N_f/2}$}\,,\\
\label{eq: U/U}
\text{$N_f$ scalars + $U(N)_{k,k+N}$} &\longleftrightarrow& \text{$N_f$ fermions + $U(k)_{-N+N_f/2,-N-k+N_f/2}$}\,,
\eea
where all matter is in the {\it fundamental} representation of the gauge group. These take the shape of level-rank dualities, but a crucial difference is that they relate critical theories of matter coupled to Chern-Simons gauge fields rather than gapped TQFTs. Across these dualities, baryons of the $SU(k)_{-N}$ theories are mapped to monopoles of the $U(N)_k$ theories. We list our conventions for the non-Abelian Chern-Simons gauge fields in the Appendix. 

Using these dualities as building blocks, it is possible to obtain new dualities relating the Abelian \textbf{Theory A} to a non-Abelian \textbf{Theory B}. The dualities obtained in this section are described in Refs. \cite{Hsin2016,Aharony2016a}, although we show in Section \ref{sec: generalized NASS} that new, more general dualities can be obtained with an analogous strategy. To begin, let us consider the case of a single layer $k=1$ of bosons at $\nu=1/2$. The Landau-Ginzburg theory for this state consists of Wilson-Fisher bosons $\Phi$ coupled to a $U(1)_2$ gauge field $a$, 
\be
\label{eq: U(1)2}
\mathcal{L}_A=|D_a\Phi|^2-|\Phi|^4+\frac{2}{4\pi}ada+\frac{1}{2\pi}Ada\,.
\ee
We start by invoking an Abelian boson-fermion duality, Eq. \eqref{eq: SU/U} with $N=k=1$, which relates a Wilson-Fisher boson to a Dirac fermion with a unit of flux attached \cite{Seiberg2016,Karch2016},
\be
|D_A\Phi|^2-|\Phi|^4\longleftrightarrow i\bar\psi\slashed{D}_b\psi-\frac{1}{2}\frac{1}{4\pi}bdb+\frac{1}{2\pi}bdA-\frac{1}{4\pi}AdA\,, \label{eqn:fermionization}
\ee
where $b$ is a new dynamical $U(1)$ gauge field\footnote{Throughout this paper, we approximate the Atiyah-Patodi-Singer $\eta$-invariant by a level-$1/2$ Chern-Simons term and include it in the action.}
. Applying this duality to $\mathcal{L}_A$ by treating $a$ as a background field, one obtains \textbf{Theory C},
\be
\mathcal{L}_A\longleftrightarrow \mathcal{L}_C= i\bar\psi\slashed{D}_b\psi-\frac{1}{2}\frac{1}{4\pi}bdb+\frac{1}{4\pi}ada+\frac{1}{2\pi}ad(b+A)\,.
\ee
We can integrate out $a$ without violating the Dirac quantization condition: its equation of motion is simply $-da=db+dA$. Thus,
\be
\label{eq: duality A-C}
\mathcal{L}_A\longleftrightarrow \mathcal{L}_C=i\bar\psi\slashed{D}_b\psi-\frac{3}{2}\frac{1}{4\pi}bdb-\frac{1}{2\pi}bdA-\frac{1}{4\pi}AdA\,.
\ee
\textbf{Theory C} was motivated as a description of the $\nu=1/2$ FQH-insulator transition in Ref. \cite{Barkeshli-McGreevy-2014}. The duality \eqref{eq: duality A-C} is a special case of more general Abelian dualities described (and derived) in Refs. \cite{Mross2017,Goldman2018}. However, of those dualities, it is one of the unique ones for which the Chern-Simons level is properly quantized. Notice also that this is the duality \eqref{eq: U/U} with $N_f=N=k=1$. The reason that we took a detour through the Abelian duality will become apparent in Section \ref{sec: generalized NASS}.

Applying the duality of Eq. \eqref{eq: SU/U} to \textbf{Theory C}, we obtain \textbf{Theory B}, which consists of bosons $\phi$ coupled to a $SU(2)_1$ gauge field $u$,
\be
\label{eq: U(1)2 duality}
\mathcal{L}_A\longleftrightarrow\mathcal{L}_B=|D_{u-A\mathbf{1}/2}\phi|^2-|\phi|^4+\frac{1}{4\pi}\Tr\left[udu-\frac{2i}{3}u^3\right]-\frac{1}{2}\frac{1}{4\pi}AdA\,,
\ee
where $\mathbf{1}$ denotes the $2\times 2$ identity matrix. Like its Abelian dual, Eq. \eqref{eq: U(1)2}, this theory describes a quantum phase transition between a $\nu=1/2$ bosonic Laughlin state (gapped $\phi$ -- the topological sector is decoupled) and a trivial insulator (condensed $\phi$). Across this duality, the monopole current of \textbf{Theory A} is related to the baryon number current of \textbf{Theory B}, 
\be
\label{eq: current dictionary}
\frac{\delta\mathcal{L}_A}{\delta A}=\frac{da}{2\pi}\longleftrightarrow \frac{\delta\mathcal{L}_B}{\delta A}= -\frac{i}{2}\,  \phi^\dagger\, \overleftrightarrow{D}_{u-A\mathbf{1}/2}\,\phi-\frac{1}{2}\frac{dA}{2\pi}
\ee
Both of these currents correspond to the physical EM charge current $J_e$. We have suppressed Lorentz indices for clarity.

We can check explicitly that the $\nu=1/2$ state has particle-hole symmetry in the composite vortex variables of \textbf{Theory B}. The physical EM charge density corresponds to the zeroth component of the currents \eqref{eq: current dictionary},
\be
\rho_e=\langle J_e^0\rangle=-\frac{1}{2}\,\rho_\phi-\frac{1}{2}\frac{B}{2\pi}\,,
\ee
where $\rho_\phi$ denotes the number density of the non-Abelian composite vortices, so, when $\rho_\phi=0$, the filling fraction is
\be
\nu=-2\pi\frac{\rho_e}{B}=\frac{1}{2}\,.
\ee
This means that the $\nu=1/2$ bosonic Laughlin state can be thought of as a gapped, particle-hole symmetric phase of non-Abelian composite vortices just as well as Abelian ones! By copying this duality $k$ times, we will see in the next subsection how to obtain a non-Abelian LG theory of the RR states.

By applying the duality of Eq. \eqref{eq: U/SU} with $N=1$ and $k=2$ to \textbf{Theory A}, it is also possible to obtain a non-Abelian fermionic \textbf{Theory D} with gauge group $SU(2)_{-1/2}$. However, in this work we focus on the non-Abelian bosonic LG theories, since in these theories the nature of the topological order and anyon content are manifest. Understanding the emergence of the RR states and other non-Abelian FQH states from the perspective of these non-Abelian composite  fermion theories will be the subject of a forthcoming work. Combining all of these dualities, we see that
\begin{align}
\textbf{Theory A: } \text{a scalar}+U(1)_{2}&\longleftrightarrow\textbf{Theory D: }\text{a fermion}+SU(2)_{-1/2}\nonumber\\
\label{eq: SU(2) dualities}
\updownarrow\hspace{1.2cm}&\\
\textbf{Theory C: }\text{a fermion}+U(1)_{-3/2}&\longleftrightarrow\textbf{Theory B: }\text{a scalar}+SU(2)_1\,.\nonumber
\end{align}
It is a miracle of arithmetic that, like the boson/fermion dualities, the boson/boson and fermion/fermion dualities above also have the flavor of level-rank dualities. Indeed, it is easy to show that the topological phases of these theories are all dual to one another \cite{Hsin2016}. This can be thought of as a consequence of the fact that 
 we were able to integrate out the gauge field $a$ above without violating flux quantization. 
It is an interesting question to ask whether there are more general dualities which exhibit the same miracle. We will show that this is indeed the case in Section \ref{sec: generalized NASS}. We finally note that the dualities of Eq. \eqref{eq: SU(2) dualities} also have the feature of hosting an emergent $SO(3)$ global symmetry, a consequence of the fact $SU(2)\simeq USp(2)$ \cite{Benini2017,Cordova2017}. This symmetry is manifest upon rewriting the theory in the $USp(2)$ language, which involves replacing the single complex matter field with two (pseudo)real ones \cite{Chatterjee2019}. 



\subsection{Building Non-Abelian States from Clustering} \label{subsec:RR clustering}

Equipped with the duality \eqref{eq: U(1)2 duality}, we now revisit the construction of Ref. \cite{Fradkin1999}, which we described in Section \ref{sec: Early LG}. We again start by considering the case where \textbf{Theory A} consists of $k=2$ layers of $U(1)_2$ LG theories, 
\be
\label{eq: MR Theory A}
\mathcal{L}_A=\sum_n\left(|D_{a_{n}}\Phi_{n}|^2-|\Phi_{n}|^4\right)+\frac{2}{4\pi}a_{n}da_{n}+\frac{1}{2\pi}Ad(a_1+a_2)\,,I=1,2\,\,.
\ee
Invoking Eq. \eqref{eq: U(1)2 duality}, \textbf{Theory B} is two $SU(2)_1$ theories, 
\be
\label{eq: MR Theory B}
\mathcal{L}_{B}=\sum_{n}\left(|D_{u_n-A\mathbf{1}/2}\phi_n|^2-|\phi_n|^4\right)+\frac{1}{4\pi}\sum_n\Tr\left[u_ndu_n-\frac{2i}{3}u_n^3\right]-\frac{1}{4\pi}AdA\,,
\ee
The half-filling condition here is simply $\nu=1$. Notice that the background gauge field $A$ couples to the ``baryon number'' current of the $\phi$'s in a {\it gauge invariant} way, in contrast to the theory of Ref. \cite{Fradkin1999}. This also means that the physical bosons can be interpreted as baryons, or color singlet bound states of two $\phi$'s. However, these are monopoles from the point of view of \textbf{Theory A}.

To obtain a $SU(2)_2$ bosonic Moore-Read state 
at $\nu=1$, we again seek the symmetry breaking pattern
\be
\label{eq: MR SB pattern}
SU(2)_1\times SU(2)_1\rightarrow SU(2)_2\,.
\ee 
As described in Section \ref{sec: Early LG}, the authors of Ref. \cite{Fradkin1999} achieved this via pairing of adjoint fields so that the theory would support vortices of the order parameter with non-Abelian statistics. Instead, we will argue that singlet pairing of our \textit{fundamental} composite vortices is sufficient to both obtain this symmetry breaking pattern and to capture the full anyon spectrum from the matter content. Nevertheless, it is still possible to obtain an analogue of the theory described in Section \ref{sec: Early LG} by ``quartetting'' the composite vortices. In this case, the order parameter contributes non-trivial vortex excitations which possess non-Abelian statistics. These vortices arise because the order parameter sees $SO(3)$ rather than $SU(2)$ gauge fields, as in Ref. \cite{Fradkin1999}, and the resulting topological order again does not quite match that of the RR states. 
We provide a brief account of the quartetted phase at the end of this section.

\subsubsection{Singlet Pairing}

We pair the non-Abelian composite vortices by adding to \textbf{Theory B}, Eq. \eqref{eq: MR Theory B}, an interaction with an electromagnetically neutral fluctuating scalar field $\Sigma_{mn}(x)$,
\begin{align}
\mathcal{L}&=\mathcal{L}_B+\mathcal{L}_{\Sigma}+\mathcal{L}_{\textrm{singlet pair}}\,,\label{eq: Sigma Lagrangian 1}\\
\mathcal{L}_{\Sigma}&=\sum_{m,n}|\pd\Sigma_{mn}-iu_m\Sigma_{mn}+i\Sigma_{mn}u_n|^2-V[\Sigma]\,,\label{eq: Sigma Lagrangian 2}\\ 
\mathcal{L}_{\textrm{singlet pair}}&= -\sum_{m,n} \phi_m^\dagger \Sigma_{mn} \phi_n\,,
\end{align}
where $\Sigma_{mn}$ is Hermitian in the layer indices $m,n$, and $V[\Sigma]$ is the potential for $\Sigma$. The off-diagonal components, $\Sigma_{12}=\Sigma_{21}^\dagger$, induce interlayer pairing, while the diagonal components, $\Sigma_{11}$ and $\Sigma_{22}$, induce intralayer pairing. Under a gauge transformation, $\Sigma_{nn}$ (no summation intended) transforms in the adjoint representation of the $SU(2)$ gauge group on layer $n$, while $\Sigma_{12}$ 
transforms as a bifundamental field under the bilayer $SU(2)\times SU(2)$ gauge group,
\begin{align}
\label{eq: Sigma transform}
\Sigma_{mn} \mapsto U_m \Sigma_{mn} U_n^\dagger, \quad U_m \in SU(2)\text{ on layer $m$.}
\end{align}
In both Eq. \eqref{eq: Sigma Lagrangian 2} and Eq. \eqref{eq: Sigma transform}, left (right) multiplication indicates contraction with $\Sigma$'s color indices in the fundamental (antifundamental) representation of $SU(2)$.

In order to achieve the symmetry breaking pattern \eqref{eq: MR SB pattern}, we choose the potential $V$ so that $\Sigma_{mn}$ condenses in such a way that $\langle\phi^\dagger_1\phi_2\rangle\neq0$ while $\langle\phi_1\rangle=\langle\phi_2\rangle=0$. Explicitly, 
\begin{align}
\langle\Sigma_{nm}\rangle=M_{nm}\mathbf{1},&\qquad\,M_{11},M_{22},\det M>0
\end{align}
The requirement 
$M_{11},M_{22},\det M>0$ guarantees that the resulting effective potential for $\phi_{1,2}$ is minimized only for $\langle\phi_1\rangle=\langle\phi_2\rangle=0$, while the off-diagonal components $M_{12}=M_{21}^\dagger$ break the $SU(2)\times SU(2)$ gauge symmetry down to the diagonal $SU(2)$. As described in Section \ref{sec: Early LG}, in the low energy limit, this sets $u_1=u_2$, and the Chern-Simons levels add to yield the correct $SU(2)_2$ Chern-Simons theory (the bosonic Moore-Read state) as the low energy TQFT. 

Having obtained the $SU(2)_2$ RR state, we now show that its anyon spectrum is furnished by the non-Abelian composite vortices $\phi_{1,2}$. 
Both $\phi_1$ and $\phi_2$ carry electric charge $Q = \frac{1}{2}$ and transform in the spin-$\frac{1}{2}$ representation of the $SU(2)_2$ gauge group, endowing them with non-Abelian braiding statistics. 
These are precisely the properties of the minimal charge anyon in the $\nu=1$ bosonic Moore-Read state, the half-vortex! 
Even though there are two bosonic fields $\phi_{1,2}$, these do not represent distinct anyons: $\phi_1$ and $\phi_2$ can be freely transformed into one another via the bilinear condensate $\langle\phi_1^\dagger \phi_2\rangle$. In other words, their currents are no longer individually conserved, and the layer index is no longer a good quantum number. 
The remainder of the anyon spectrum is obtained by constructing composite operators of the $\phi$ fields or, equivalently, by fusing multiple minimal charge anyons. In the present case, the only remaining anyon is the Majorana fermion, which transforms in the spin-1 representation of $SU(2)$, and so is represented by the local bilinear $\chi_n^a=\phi^\dagger_n t^a \phi_n$ (see Table \ref{tab:MR-anyons}). 
We note that, unlike in Ref. \cite{Fradkin1999}, there are no non-trivial vortices in this approach, since an order parameter valued on $[SU(2)\times SU(2)]/SU(2)$ cannot host non-trivial vortices.

\begin{table}
 \centering
 \large
 \caption{List of quasi-particles in the $\nu=1$ bosonic Moore-Read state, their spin, $\theta$, $U(1)_{EM}$ charges, $Q$, and the corresponding operator in our LG theory. We label the anyons by the corresponding operators in the edge CFT (see e.g. Refs. \cite{Moore1991,Fradkin-2013}). Note that we do not sum over the layer index $n$.}
 \begin{tabular}{||c | c | c | c||} 
 \hline
 & $1$ (vacuum) & $\sigma e^{i\varphi/2}$ (half-vortex) & $\chi$ (Majorana fermion)\\ %
 \hline\hline
 $\theta$ & $0$ & $\frac{3}{16}$ & $\frac{1}{2}$\\ 
 \hline
 $Q$ 	  & $0$ & $\frac{1}{2}$ & $0$ \\
 \hline
 $\text{Field theory}$ & - & $\phi_n$ & $\phi^\dagger_n t^a \phi_n$ \\
 \hline
\end{tabular} \label{tab:MR-anyons}
\end{table}

The reader might object to our identification of the individual particles making up the pairs with the fundamental anyons, since the energy cost to break up a pair 
will be on the order of the UV cutoff. However, this is not a significant shortcoming of our construction, since anyons are only well defined upon projecting into the (topologically ordered) ground state. They should therefore always be viewed as infinite energy excitations representated as Wilson lines. 


\subsubsection{Quartetting and Vortices \label{sec:Read Rezayi Quartetting}}

Although singlet pairing is sufficient to obtain the RR states,  
it is interesting to consider an alternative mechanism for obtaining non-Abelian states that more closely resembles the construction of Ref. \cite{Fradkin1999} that was discussed in Section \ref{sec: Early LG}. In this scenario, rather than pairing the non-Abelian bosons of \textbf{Theory B} \eqref{eq: MR Theory B}, we imagine {\it quartetting} them. 
To do this, we define the adjoint operators,
\be
B^a_n=\phi^\dagger_n t^a\phi_n\,,
\ee
where 
the repeated $n$ index on the right hand side is not summed over. These operators are neutral under $U(1)_{\mathrm{EM}}$, and they will serve the same purpose for us here as the $B^a_n$ fields disucussed in Section \ref{sec: review} and Ref. \cite{Fradkin1999}. We thus consider a pairing interaction of the $B^a_n$'s, or a {\it quartetting} interaction of the $\phi$'s, by introducing a scalar field $\mathcal{O}$ to mediate the pairing interaction,
\be
\mathcal{L}_{\mathrm{quartet}}=\lambda\,B^a_1\,\mathcal{O}^{ab}\,B^b_2=\lambda\,(\phi^\dagger_1 t^a\phi_1)\,\mathcal{O}^{ab}\,(\phi^\dagger_2 t^b\phi_2)\,.
\ee
The quartetted phase, where $\langle\mathcal{O}^{ab}\rangle=v\delta^{ab}$ and $\langle\phi_1\rangle=\langle\phi_2\rangle=0$, is accessed by adding a suitable potential $V[\mathcal{O}]$ and ensuring that $\phi_{1,2}$ are gapped via a mass term $-m^2\sum_n|\phi_n|^2$. Because $\mathcal{O}$ radiatively acquires a kinetic term of the form of a gauged nonlinear sigma model (NLSM), the resulting effective theory in the quartetted phase is
\be
\label{eq: new NLSM}
\mathcal{L}_{\mathrm{eff}}=\mathcal{L}_B+\mathcal{L}_{\textrm{quartet}} - m^2\sum_n|\phi_n|^2-V[\mathcal{O}]+ \kappa\,\Tr\left[\mathcal{O}^{-1}D_{u_1-u_2}\mathcal{O}\,\mathcal{O}^{-1}D_{u_1-u_2}\mathcal{O}\,\right] 
\ee
where $\kappa$ is a coupling constant defined so that $\mathcal{O}$ is properly normalized.

Since $\mathcal{O}$ transforms in the adjoint representation of the $SU(2)$ of each layer, it is blind to their $\mathbb{Z}_2$ centers. This means that the quartetted phase hosts not only the non-Abelian $SU(2)_2$ topological order (since $u_1-u_2$ is again Higgsed), but also an additional Abelian $\mathbb{Z}_2$ sector. Explicitly, as noted in Section \ref{sec: Early LG}, the condensation of $\mathcal{O}$ yields the symmetry breaking pattern $SU(2) \times SU(2) \to SU(2)_{\mathrm{diag}} \times \mathbb{Z}_2$, where the residual $\mathbb{Z}_2$ can be chosen to act on either $\phi_1$ or $\phi_2$ (amounting to a choice of basis). 
Hence, the full topological order of the ground state is $SU(2)_2\times \mathbb{Z}_2$. This is also true of the original construction of Ref. \cite{Fradkin1999}, meaning that the singlet pairing mechanism discussed above carries the significant advantage that it yields the $\nu=1$ Moore-Read state {\it alone}, with no additional Abelian sector. We therefore focus on singlet pairing for the remainder of this work. 

How do we account for the new Abelian anyon content? As discussed in Section \ref{sec: Early LG}, because of the order parameter's blindness to the $\mathbb{Z}_2$ centers, the NLSM above admits vortex solutions. 
These vortices can carry fluxes of {\it both} of the residual $\mathbb{Z}_2$ and $SU(2)$ gauge groups, and so they possess non-trivial braiding statistics with respect to each other and the scalar fields. However, since the $B^a_n$ fields here are electrically neutral, the vortices of the order parameter should not carry any electric charge either. These vortices should therefore correspond to anyon excitations which are distinct from those that can be obtained from the $\phi_{1,2}$ fields alone, as these fields carry electric charge. We leave a detailed understanding of this Abelian sector to future work. 

As in the singlet pairing case, this quartetting procedure can be generalized to the case of $k$ layers, or $\nu=k/2$, which can be easily shown to have $SU(2)_k\times\mathbb{Z}_2^{k-1}$ topological order (each factor of $\mathbb{Z}_2$ corresponds to the unbroken center of a broken $SU(2)$). In the next subsection, we describe how both the singlet pairing and quartetting constructions can be generalized to the remaining RR fillings through a flux attachment transformation.

\subsection{Generating the Full Read-Rezayi Sequence through Flux Attachment}
\label{subsec: flux attachment}

By attaching $M$ fluxes to the $k$-layer generalization of \textbf{Theory A} \eqref{eq: MR Theory A} and performing the same transformation on \textbf{Theory B} \eqref{eq: MR Theory B}, it is possible to obtain LG theories of the remaining RR states at filling fractions
\be
\label{eq: RR filling}
\nu=\frac{k}{Mk+2}\,.
\ee
Flux attachment can be performed on \textbf{Theory A} as a modular transformation $\mathcal{S}\mathcal{T}^M\mathcal{S}$ \cite{Kivelson1992,Witten2003}, where
\be
\mathcal{S}: \mathcal{L}[A]\mapsto\mathcal{L}[b]+\frac{1}{2\pi}Adb\,,\qquad\mathcal{T}:\mathcal{L}[A]\mapsto\mathcal{L}[A]+\frac{1}{4\pi}AdA\,,
\ee
where again $A$ is the background EM field, and $b$ is a new dynamical $U(1)$ gauge field. Thus, attaching $M$ fluxes to \textbf{Theory A} amounts to 
\be
\mathcal{S}\mathcal{T}^M\mathcal{S}:\mathcal{L}_A[A]\mapsto\mathcal{L}_A[b]+\frac{1}{2\pi}cd(b+A)+\frac{M}{4\pi}cdc\,,
\ee  
where $c$ is a new dynamical $U(1)$ gauge field. It is straightforward to see that this transformation is equivalent to the usual attachment of $M$ fluxes to the composite bosons (related to the composite vortex variables -- or Laughlin quasiparticles -- of \textbf{Theory A} by boson-vortex duality \cite{Peskin1978,Dasgupta1981}). One of the insights of Refs. \cite{Seiberg2016,Karch2016} was that the modular group PSL$(2,\mathbb{Z})$ generated by $\mathcal{S}$ and $\mathcal{T}$ can generate new dualities from old ones. Restricting for the moment to $k=2$ layers, the transformed \textbf{Theory A} is dual to 
\be
\tilde{\mathcal{L}}_{B}=\sum_n\left(|D_{u_n-b\mathbf{1}/2}\phi_n|^2-|\phi_n|^4\right)+\frac{1}{4\pi}\sum_n\Tr\left[u_ndu_n-\frac{2i}{3}u_n^3\right]-\frac{1}{4\pi}bdb+\frac{1}{2\pi}cd(b+A)+\frac{M}{4\pi}cdc\,.
\ee 
We can repackage the $SU(2)$ gauge fields $u_n$ as new $U(2)$ gauge fields $u'_n$ with trace $\Tr[u'_1]=\Tr[u'_2]=b$. This gluing of the traces together can be implemented by introducing a new auxiliary gauge field $\alpha$,
\bea
\tilde{\mathcal{L}}_B=&&\sum_n\left(|D_{u'_n}\phi_n|^2-|\phi_n|^4\right)+\frac{1}{4\pi}\sum_n\Tr\left[u'_ndu'_n-\frac{2i}{3}u_n^{'3}\right]\\
&&-\frac{2}{4\pi}\Tr[u'_1]d\Tr[u'_1]+\frac{1}{2\pi}cd(\Tr[u'_1]+A)+\frac{M}{4\pi}cdc+\frac{1}{2\pi}\alpha d\left(\Tr[u_1']-\Tr[u_2']\right)\,.\nonumber
\eea
This transformation does not impact the singlet pairing nor the quartetting procedure discussed in the previous subsection, and it readily generalizes to $k$ layers (more constraints need to be introduced in that case to glue the Abelian gauge fields together). We therefore obtain the $SU(2)_2$ Chern-Simons theory at low energies, albeit with the additional Abelian sector introduced above. 
For the general case of $k$ layers, 
the $u'_n$'s on each layer are set equal to one another, and the low energy TQFT is a $U(2)_{k,-2k}\times U(1)_M$ Chern-Simons-BF theory given by
\be
\mathcal{L}=\frac{k}{4\pi}\Tr\left[u'du'-\frac{2i}{3}u^{'3}\right]-\frac{k}{4\pi}\Tr[u']d\Tr[u']+\frac{1}{2\pi}cd(\Tr[u']+A)+\frac{M}{4\pi}cdc\,.
\ee
This is indeed the proper bulk TQFT describing the RR states at filling \eqref{eq: RR filling}, first described in Ref. \cite{Seiberg2016a}. 
As in the case of the $\nu=1$ bosonic Moore-Read state discussed above, the fundamental scalars (i.e. the composite vortices) comprise the minimal charge anyons, here possessing electric charge $Q = 1/(Mk+2)$. This is the expected result for the minimal charge anyon in the general RR states.


\section{Generalization to Non-Abelian $SU(N_f)$-Singlet States}
\label{sec: generalized NASS}
Having derived a LG theory for the RR states, we will now demonstrate how our construction can be naturally extended to 
the generalized non-Abelian $SU(N_f)$-singlet states occuring at fillings
\be
\label{eq: NASS fillings 2}
\nu=\frac{k N_f}{N_f+1+kMN_f}\,,\qquad k,N_f,M\in\mathbb{Z}\,.
\ee
These are clustered states in which $k$ represents the number of local particles (fermions or bosons for odd and even $M$, respectively) in a cluster, $M$ the number of attached Abelian fluxes, and $N_f$ the number of internal degrees of freedom. Like the RR states, which correspond to $N_f=1$, we will show that these states can also be obtained by pairing starting from a parent multi-layer Abelian LG theory. 
The particular Abelian states we will target are the $N_f$-component generalizations of the Halperin $(2,2,1)$ states.
In parallel to Section \ref{section: Read Rezayi Bosonization}, we will show that the LG theories of these Abelian states satisfy a new non-Abelian bosonization duality. This duality relates the Abelian LG theory of the generalized Halperin states to an $SU(N_f+1)_1$ Chern-Simons-matter theory. 
That this is possible is perhaps not surprising given that the $N_f$-component $(2,2,1)$ state is known to have an edge theory which furnishes a representation of the $SU(N_f+1)_1$ Kac-Moody algebra, as we shall review below \cite{Ardonne1999,Read1990,Frohlich1991,Fuji2017}. The generalized NASS states are then obtained by singlet pairing of the dual non-Abelian bosons. 

\subsection{Motivation: ``Projecting Down'' to the Generalized NASS States}
Just as the RR states are naturally understood starting with the $\nu=1/2$ Laughlin state by way of ``projecting down,'' the generalized NASS states can be built up from $N_f$-component generalizations of the Halperin $(2,2,1)$ spin-singlet state \cite{Halperin1983}. These (bosonic) states are {\it Abelian} and correspond to $M=0,k=1$. 
These states are described by the wave functions
\begin{align}
\Psi^{(221)}_{N_f}(\{z_{i}^\sigma\}) = \prod_{\sigma=1}^{N_f} \prod_{i<j} (z_{i}^\sigma - z_{j}^\sigma )^{2} \prod_{\sigma < \sigma'}^{N_f} \prod_{i,j} (z_{i}^\sigma - z_{j}^{\sigma'} )^{1}  e^{-\frac{1}{4}\sum_{\sigma,i} |z^\sigma_i|^2}, \label{eqn:halperin-wavefn}
\end{align}
where $z_i^\sigma = x_i^\sigma + i y_i^\sigma$ denotes the complex coordinates of the $i^\mathrm{th}$ boson with component index $\sigma$. In direct analogy with the $\nu=k/2$ RR states, the generalized NASS wave functions for general $k$ (but still $M=0$) may be obtained by symmetrizing over a product of $k$ copies of the $N_f$-component $(2,2,1)$ wave function \cite{Sterdyniak2013},
\begin{align}
\Psi_{k,N_f} = \mathcal{S}_k \left[ \prod_{i=0}^{k-1} \Psi^{(221)}_{N_f}(z_{1+iN/k},\dots,z_{(i+1)N/k})\right]. \label{eqn:nass-wavefn}
\end{align}
where the symmetrization operation $\mathcal{S}_k$ is morally the same as the one defined in Section \ref{sec: RR wave fncs}. Again, the form of the wave function makes explicit the clustering of bosons characteristic of non-Abelian states. The wave functions for general $M$ are obtained by multiplying 
$\Psi_{k,N_f}$ by a $\nu=\frac{1}{M}$ Laughlin factor. Note that setting $N_f=1$ recovers the RR wave functions \eqref{eqn:RR-wavefn}. 

The 
generalized NASS wave functions \eqref{eqn:nass-wavefn} should also be expresssible as correlators of the $SU(N_f+1)_k$ WZW CFT for $M=0$ and of the $[U(1)]^{N_f}\times SU(N_f+1) / [U(1)]^{N_f}$ coset CFT for $M>0$. Although this appears to have only been discussed explicitly for $N_f=1,2,3$ \cite{Cappelli2001,Schoutens2002,Reijnders2002}, we will assume that this holds true for general $N_f$. 
We thus expect the corresponding bulk theories for the generalized NASS states to be
$SU(N_f+1)_k$ Chern-Simons theories. 

For the $N_f$-component Halperin states ($k=1$), the presence of this ``hidden'' $SU(N_f+1)$ representation can be motivated as follows. These states are described by a $N_f\times N_f$ $K$-matrix and $N_f$-component charge vector $q$,
\begin{align}
K = \begin{pmatrix}
2      & 1      & 1 & \dots  & 1 & 1 \\
1      & 2      & 1 & \dots  & 1 & 1 \\
1      & 1      & 2 &        &   & 1 \\
\vdots & \vdots &   & \ddots &   & \vdots \\
1      & 1      &   &        & 2 & 1 \\
1      & 1      & 1 & \dots  & 1 & 2
\end{pmatrix}, \qquad
q = \begin{pmatrix}
1 \\
\vdots \\
1
\end{pmatrix}. \label{eqn:halperin-221-kmatrix}
\end{align}
The form of the charge vector reflects the fact that the physical bosonic excitations of each species each carry the same EM charge, and it can read off that the Hall conductivity is $\sigma_{xy} = q^T K^{-1} q {e^2\over h} =  {N_f \over N_f+1}{e^2\over h}$. Under a particular change of basis $\tilde{K} = G^T K G$ and $\tilde{q} = Gq$, $G \in SL(N_f,\mathbb{Z})$, 
$K$ can be shown to be related to the Cartan matrix of $SU(N_f+1)$ \cite{Frohlich1991,Fuji2017}, 
\begin{align}
G = \begin{pmatrix}
1 & -1 &     &    & \\
  &  1 & -1   &    &  \\
  &  & \ddots & \ddots &       \\
  &  &  & 1 &  -1     &  \\
  &  &        &       &   1  \\
\end{pmatrix} \Rightarrow 
\tilde{K} = \begin{pmatrix}
2      & -1      & 0 & \dots  & 0 & 0 \\
-1      & 2      & -1 & \dots  & 0 & 0 \\
0      & -1      & 2 &        &   & 0 \\
\vdots & \vdots &   & \ddots &   & \vdots \\
0      & 0      &   &        & 2 & -1 \\
0      & 0      & 0 & \dots  & -1 & 2
\end{pmatrix}, \qquad
\tilde{q} = \begin{pmatrix}
0 \\
0 \\
\vdots \\
1
\end{pmatrix}. \label{eqn:cartan-k-matrix}
\end{align}
Using this fact, one can show that the 
edge theory 
defined by $\tilde{K}$ supports a $SU(N_f+1)_1$ Kac-Moody algebra (see e.g. Refs. \cite{Difrancesco-1997, Fuji2017} for a derivation), and hence is equivalent to the $SU(N_f+1)$ WZW CFT. 
Consequently, the corresponding bulk theory of the $N_f$-component $(2,2,1)$ Halperin state is a $SU(N_f+1)_1$ Chern-Simons theory. This is the $N_f$-component generalization of the level-rank duality $U(1)_2\leftrightarrow SU(2)_1$ described in Section \ref{sec: review}. 

This discussion indicates that we should expect the LG theories of the generalized NASS states can be obtained from pairing $k$ copies of the $N_f$-component $(2,2,1)$ Halperin state. Because this state is level-rank dual to a $SU(N_f+1)_1$ theory, we might expect that there is a non-Abelian Chern-Simons-matter theory  duality also taking this shape, from which we can build a LG theory of the non-Abelian states. We now show that this is indeed the case. 

\subsection{Non-Abelian Duals of $N_f$-Component Halperin $(2,2,1)$ States}
The necessary non-Abelian duality can be constructed by starting with the Abelian LG theory for the $N_f$-component Halperin state, which we again call \textbf{Theory A}. This theory consists of $N_f$ species of Wilson-Fisher bosons $\Phi_{I}$, $I=1, \ldots, N_f$, each coupled to a $U(1)$ Chern-Simons gauge fields $a_{I}$, 
\begin{align}
\mathcal{L}_A=\sum_{I=1}^{N_f} \left(|D_{a_{I}} \Phi_{I}|^2-|\Phi_{I}|^4\right)+\frac{1}{4\pi} \sum_{I,J=1}^{N_f} K_{IJ} a_{I} da_{J}+\frac{1}{2\pi} \sum_{I=1}^{N_f} q_I Ada_{I},
\label{eqn:k-matrix-theory}
\end{align}
where $K$ and $q$ are given in Eq. \eqref{eqn:halperin-221-kmatrix}. The $N_f$-component Halperin state corresponds to the phase in which all of the $\Phi_I$ fields -- the Laughlin quasiparticles -- are gapped. We emphasize that there is no continuous $SU(N_f)$ global symmetry rotating the $\Phi_I$ fields manifest in \textbf{Theory A}. Instead, there is only a discrete exchange symmetry of the $\Phi_I$ fields. 

Following the reasoning laid out in Section \ref{subsec:numerology}, 
we now show that this theory is dual to one of $N_f$ Wilson-Fisher bosons coupled to a {\it single} $SU(N_f+1)$ gauge field. Similar  dualities have also been described in Ref. \cite{Karch2016a}.
We start by applying the Abelian boson-fermion duality of Eq. \eqref{eqn:fermionization} to each scalar $\Phi_I$, treating the $a_I$'s as background fields, to obtain the Dirac fermion \textbf{Theory C}, 
\begin{align}
\begin{split}
\mathcal{L}_A \longleftrightarrow \mathcal{L}_C = &\sum_{I=1}^{N_f} i\bar{\psi}_I \slashed{D}_{b_I} \psi_I + \sum_{I=1}^{N_f} \frac{1}{4\pi} a_I da_I + \sum_{I=1}^{N_f} \sum_{J=I+1}^{N_f} \frac{1}{2\pi} a_I da_J + \sum_{I=1}^{N_f} \frac{1}{2\pi} A d a_I \\
&+ \sum_{I=1}^{N_f} \left[ -\frac{1}{2}\frac{1}{4\pi} b_I d b_I + \frac{1}{2\pi} b_I d a_I \right].
\end{split}
\end{align}
As in the example discussed in Section \ref{subsec:numerology}, the $a_I$ fields can be safely integrated out while respecting the Dirac flux quantization condition. This is because all of the Chern-Simons terms have coefficient equal to unity. On integrating out one of the $a_I$ fields, 
the remaining 
ones become Lagrange multipliers enforcing the constraints $b_I = b_1 \equiv b$. Integrating out the remaining $a_I$'s, we find that \textbf{Theory C} can be rewritten as one of fermions coupled to a single dynamical gauge field,
\begin{align}
\mathcal{L}_C =  \sum_{I=1}^{N_f} i\bar{\psi}_I \slashed{D}_{b} \psi_I- \frac{N_f + 2}{2}\frac{1}{4\pi} b d b - \frac{1}{2\pi} b dA - \frac{1}{4\pi} A d A.
\end{align}
In contrast to \textbf{Theory A}, \textbf{Theory C} has a manifest $SU(N_f)$ global flavor symmetry\footnote{See Ref. \cite{Benini2017} for a more detailed discussion of global symmetries in non-Abelian dualities.} since the fermions all couple in the same way to the gauge field $b$. This symmetry is thus an {\it emergent} symmetry from the point of view of \textbf{Theory A}. 

We may now apply the non-Abelian duality \eqref{eq: SU/U} to \textbf{Theory C}, leading to a non-Abelian bosonic \textbf{Theory B},
\begin{align}
\mathcal{L}_B &= \sum_{I=1}^{N_f}|D_{u - \frac{1}{N_f + 1} A \mathbf{1}} \phi_I|^2 - |\phi|^4 + \frac{1}{4\pi} \mathrm{Tr}\left[udu - \frac{2i}{3}u^3 \right] - \frac{1}{4\pi} \frac{N_f}{N_f+1}AdA\,. 
\end{align}
where $-|\phi|^4$ denotes tuning to the Wilson-Fisher fixed point consistent with a global $SU(N_f)$ symmetry. We will again refer to the $\phi_I$ fields as the non-Abelian composite vortices. It will be convenient in the subsection below to re-express this theory as a $U(N_f+1)$ gauge theory with a constraint, 
\be
	\mathcal{L}_B= \sum_{I=1}^{N_f}|D_u \phi_I|^2 - |\phi|^4 + \frac{1}{4\pi} \mathrm{Tr}\left[udu - \frac{2i}{3}u^3 \right] + \frac{1}{2\pi} \alpha d(\mathrm{Tr}\left[ u \right] - A) - \frac{1}{4\pi} A d A\,,  \label{eqn:221-na-u(n)}
\ee
where we have introduced a $U(1)$ gauge field $\alpha$. We have thus obtained a {\it new} triality,
\begin{align}
&\textbf{Theory A: } N_f \,\text{ scalars} + U(1)\text{ }K\text{-matrix theory of Eq. \eqref{eqn:halperin-221-kmatrix}} \nonumber \\
&\hspace{4.2cm} \updownarrow \label{eqn:Halperin Dualities}\\
&\textbf{Theory C: }N_f \,\text{ fermions} + U(1)_{-\frac{N_f + 2}{2}}\longleftrightarrow\textbf{Theory B: }N_f \,\text{ scalars} + SU(N_f + 1)_1\,.\nonumber
\end{align}
This is the main result of this subsection. 
It is interesting that, for our particular choice of $K$-matrix in \textbf{Theory A}, we have obtained a non-Abelian dual theory in which the rank of the gauge group depends on the number of matter species and in which an emergent $SU(N_f)$ symmetry appears. 
Such trialities can be extended by applying the modular transformation $\mathcal{S}\mathcal{T}^{P-1}\mathcal{S}$ (flux attachment) to each side, transforming the $K$ matrix of \textbf{Theory A} to that of the $N_f$-component $(P+1,P+1,P)$ Halperin states. 
The family of Abelian composite fermion theories obtained by this transformation has been conjectured to describe plateau transitions in fractional Chern insulators \cite{Lee2018}. 

Notice that Eq. \eqref{eqn:Halperin Dualities} does not contain a non-Abelian fermionic theory analogous to \textbf{Theory D} in Eq. \eqref{eq: SU(2) dualities}. 
That is not to say such a theory does not exist. 
As with the RR states, we leave to future work a full inquiry into how the NASS states, to be discussed in the next section, may arise in a fermionic picture.

\subsection{Generating the Non-Abelian $SU(N_f)$-Singlet Sequence from Clustering}
With the non-Abelian composite vortex description of the $N_f$-component $(2,2,1)$ states in hand, we can 
follow the pairing procedure of 
Section \ref{subsec:RR clustering} to generate the generalized NASS sequence. Unlike in Section \ref{section: Read Rezayi Bosonization}, in this section we will consider LG theories for general $k,M$, and $N_f$ from the outset. Our \textbf{Theory A} will thus consist of $k$ layers of LG theories of the $N_f$-flavor Halperin $(2,2,1)$ states, 
\begin{equation}
\mathcal{L}_A =\sum_{I,n} \left(|D_{a_{I,n}} \Phi_{I,n}|^2-|\Phi_{I,n}|^4\right)+\frac{1}{4\pi} \sum_{I,J,n} K_{IJ} a_{I,n} da_{J,n}+\frac{1}{2\pi} \sum_{I,n} q_I Ada_{I,n}, \label{eqn:layered-k-matrix-theory}
\end{equation}
where again the $K$-matrix and charge vector are given by Eq. \eqref{eqn:halperin-221-kmatrix}, and $n=1,\dots , k$ denotes the layer index. 
Applying the duality \eqref{eqn:221-na-u(n)} to each layer, this theory is dual to the non-Abelian \textbf{Theory B},
\begin{align}
\mathcal{L}_B = \sum_{I,n}|D_{u_n}\phi_{I,n}|^2-\sum_n|\phi_{n}|^4 + \sum_{I,n} \mathcal{L}_{U(N_f+1)}[u_n] + \frac{1}{2\pi} \sum_{I,n} \alpha_n d(\mathrm{Tr}\left[ u_n \right] - A) - \frac{k}{4\pi} A d A.
\end{align}
Here, lower case Latin letters denote a layer index, upper case Latin letters a flavor index. 
We have also defined, for compactness,
\begin{align}
\mathcal{L}_{U(N)}[u] \equiv \frac{1}{4\pi} \mathrm{Tr}\left[udu - \frac{2i}{3}u^3 \right].
\end{align}
We introduce $M$ via flux attachment, or application of the modular transformation $\mathcal{S}\mathcal{T}^M\mathcal{S}$, as in Section \ref{subsec: flux attachment}. This yields a sequence of descendant theories labelled by $k,M$, and $N_f$,
\begin{align}
\begin{split}
\tilde{\mathcal{L}}_B &= \sum_{I,n}|D_{u_n}\phi_{I,n}|^2-\sum_n|\phi_{n}|^4 + \sum_{n} \mathcal{L}_{U(N_f+1)}[u_n] + \frac{1}{2\pi} \sum_{n} \alpha_n d(\mathrm{Tr}\left[ u_n \right] - a) \\
&\qquad - \frac{k}{4\pi} a d a + \frac{1}{2\pi} a db + \frac{M}{4\pi} bdb + \frac{1}{2\pi} b d A .
\end{split} \label{eq:general-halperin-theory-b}
\end{align}
We are now in a position to consider singlet pairing between the different layers. One can also consider quartetting the composite vortices, but this only leads to additional Abelian sectors, as in the RR case. 

Singlet pairing between the fundamental scalars is again mediated via a dynamical scalar field, $\Sigma_{m,n}(x)=\Sigma^\dagger_{n,m}(x)$, transforming in the bifundamental representation of the $SU(N_f+1)$ factor on layer $m$ and on layer $n$, i.e. $\Sigma_{m,n} \mapsto U_m \Sigma_{m,n} U_n^\dagger$, where $U_n,U_m\in SU(N_f+1)$. Note that the $U(1)$ gauge transformations cancel out, as the $\alpha_n$ fields force all the $U(1)$ gauge fields $\mathrm{Tr}[u_n]$ to be equal. If we require that $\Sigma_{m,n}$ be a flavor singlet, its coupling to the non-Abelian composite vortices is therefore  
\begin{align}
\mathcal{L}_{\textrm{singlet pair}} = -\sum_{m,n,I}\phi_{I,m}^\dagger\, \Sigma_{m,n}\, \phi_{I,n}\,.
\end{align}
As before, the off-diagonal terms induce inter-layer pairing, while the diagonal terms can be used to ensure that $\langle\phi_{I,n}\rangle=0$. 
Thus, 
we obtain a non-Abelian state when 
$\Sigma_{m,n}$ condenses in such a way that it enforces the constraint $u_n \equiv u'$ for all $n$. 
Putting these pieces together, we find that the paired phase is governed by the TQFT 
\begin{align}
\mathcal{L}_{\text{eff}} = k\, \mathcal{L}_{U(N_f+1)}[u'] - \frac{k}{4\pi} \mathrm{Tr}[u'] d \mathrm{Tr}[u'] + \frac{1}{2\pi} \mathrm{Tr}[u'] db + \frac{M}{4\pi} bdb + \frac{1}{2\pi} b d A \label{eqn:nass-tqft} 
\end{align}
Integrating out the fluctuating gauge fields indeed yields the correct Hall response,
\begin{align}
\sigma_{xy} = \frac{kN_f}{N_f+1+kMN_f}\frac{e^2}{h},
\end{align}
which is the expected result for the generalized NASS states. 

As in our LG theories of the RR states, the fundamental scalars $\phi_{I,n}$ 
correspond to the minimal charge anyons. Indeed, one can check from the equations of motion that the fundamental scalar fields each carry charge $Q = \frac{1}{N_f + 1+MkN_f}$, which reduces to the expected result for the minimal charge anyons of the RR and non-Abelian spin singlet states for $N_f=1$ and $N_f=2$, respectively. Additionally, in the paired phase, the condensation of the bilinears $\phi^\dagger_{I,m}\phi_{I,n} + H.c.$ (no sum on $I$) ensures that all the $\phi_{I,n}$, for fixed $I$, are indistinguishable, removing the redundancy of the layer degree of freedom. In particular, because we took the pairing interaction to be diagonal in the flavor indices, there is no mixing between flavors on different layers. Hence the fundamental scalar excitations should still transform into each other under the diagonal $SU(N_f)$ subgroup of the original $SU(N_f) \times \dots \times SU(N_f)$ global symmetry. Consequently, our theory reproduces the desired anyon spectrum, and we conclude that we have obtained a LG theory for the generalized NASS states.

\section{Discussion}
\label{sec: discussion}

Using non-Abelian boson-fermion dualities, we have presented a physical pairing mechanism by which the non-Abelian Read-Rezayi states and their generalizations, the non-Abelian $SU(N_f)$-singlet states, may be obtained by ``projecting down" from parent Abelian states. These dualities relate the usual Abelian LG theories of the parent state to theories of non-Abelian ``composite vortices," which pair to form the non-Abelian FQH state. While this pairing amounts to condensing local operators in the non-Abelian theory, this is not the case in the original Abelian LG theory of Laughlin quasiparticles, in which the composite vortices are monopoles. In the process of developing these theories, we have described a new triality \eqref{eqn:Halperin Dualities} which parallels a level-rank duality apparent from CFT/ideal wave function considerations and which has the interesting property that it involves a non-Abelian gauge theory with rank depending on the number of matter species. We believe that this approach for obtaining physically motivated bulk descriptions of non-trivial gapped phases represents a promising direction for future applications of duality to condensed matter physics which has thus far been under-explored.  

Our construction contrasts with earlier bulk descriptions of non-Abelian FQH states in important ways. The use of non-Abelian boson-fermion dualities, which relate parent quantum critical points, or Landau-Ginzburg effective field theories, provides a clear mapping to theories of non-Abelian ``composite vortex'' variables which are manifestly gauge invariant, unlike in earlier approaches that invoked level-rank duality deep in the topological phase \cite{Fradkin1999,Ardonne1999}. Additionally, we showed that these earlier approaches in fact lead to a superfluous Abelian sector on top of the desired non-Abelian topological order. The use of non-Abelian dualities also avoids the issues inherent to parton constructions \cite{Wen1991,Wen1999,Barkeshli2010}, which provide a perhaps larger class of fractionalized descriptions but rely on the assumption that the fractionalized particles are not confined. This is  in spite of the fact that they are generally charged under non-Abelian gauge fields without Chern-Simons terms and, as such, are known to be confining  in 2+1 dimensions. Consequently, it is likely that many partonic descriptions are on unstable dynamical footing. 
We anticipate that many more exotic FQH and otherwise topologically ordered states can be targeted with our approach. Again, we can draw inspiration from edge CFT and ideal wave function approaches. For instance, the spin-charge separated spin-singlet states of Ref. \cite{Ardonne2002} can both be related to a parent bilayer Abelian state and be obtained from conformal blocks of an $SO(5)$ WZW theory.
There exist, in fact, Chern-Simons-matter dualities involving precisely $SO(N)$ (and many other) gauge groups \cite{Aharony2016a,Cordova2018}, which suggests that it may be possible to formulate 
non-Abelian Landau-Ginzburg theories of these states. 
It is perhaps also possible to apply our approach to generating bulk parent descriptions of the orbifold FQH states \cite{Barkeshli2011}, 
which can involve an interesting interplay of usual gauge symmetries with gauged higher-form symmetries \cite{Kapustin2014,Gaiotto2015}. 

In this work, we have focused on understanding non-Abelian states via pairing of non-Abelian bosonic matter. However, as described in Section \ref{section: Read Rezayi Bosonization}, a non-Abelian composite fermion description is available for the $\nu=\frac{1}{2}$ Laughlin states. 
In the parent Abelian phase, these fermions feel a magnetic field and fill an integer number of Landau levels. Pairing across layers of these integer quantum Hall states appears to lead in fact to $SU(2)_{-k}$ theories, which may be connected to the particle-hole conjugates of the RR states (note the sign of $k$). One may also consider starting not from multiple layers of FQH phases but instead of the (fermionic) compressible states at filling $\nu=1/2n$, for which Dirac fermion theories have been proposed \cite{Son2015,Goldman2018a}. It is possible that applying non-Abelian dualities to these theories may provide an avenue for developing exotic non-Abelian {\it excitonic} phases. We plan to provide a general discussion of composite fermion approaches to generating non-Abelian states in future work.

We lastly comment on the possible connection of the theories presented here to numerical studies of transitions between Abelian and non-Abelian states in bilayers \cite{Crepel-2019,Zhu2016,Liu2015,Geraedts2015a,Peterson2015,Papic2010}. To the extent that these transitions are continuous, it is an exciting possibility that they are in the universality class of the quantum critical theories presented here. However, since these theories are very strongly coupled, the only analytic techniques against which this can be checked are large-$N$ approaches, which may describe a wholly different fixed point. Perhaps eventually the conformal bootstrap will be able to shed light on this issue.



\section*{Acknowledgements}

We thank M. Barkeshli, M. Mulligan, S. Raghu, and D. T. Son for discussions and comments on the manuscript. HG is supported by the National Science Foundation (NSF) Graduate Research Fellowship Program under Grant No. DGE-1144245. RS acknowledges the support of the Natural Sciences and Engineering Research Council of Canada (NSERC) [funding reference number 6799-516762-2018]. This work was also supported in part by the 
NSF under grant No. DMR-1725401 at the University of Illinois (EF).

\begin{appendix}
\section{Chern-Simons Conventions}
In this appendix, we lay out our conventions for non-Abelian Chern-Simons gauge theories. We define $U(N)$ gauge fields $a_\mu=a^b_\mu t^b$, where $t^b$ are the (Hermitian) generators of the Lie algebra of $U(N)$, which satisfy $[t^a,t^b]=if^{abc}t^c$, where $f^{abc}$ are the structure constants of $U(N)$. The generators are normalized so that $\Tr[t^bt^c]=\frac{1}{2}\delta^{bc}$. The trace of $a$ is a $U(1)$ gauge field, which we require to satisfy the Dirac quantization condition,
\be
\label{eq: flux quantization}
\int_{S^2} \frac{d\Tr[a]}{2\pi}\in\mathbb{Z}\,.
\ee
In general, the Chern-Simons levels for the $SU(N)$ and $U(1)$ components of $a$ can be different. We therefore adopt the standard notation \cite{Aharony2016},
\be
U(N)_{k,k'}=\frac{SU(N)_k\times U(1)_{Nk'}}{\mathbb{Z}_N}\,.
\ee
By taking the quotient with $\mathbb{Z}_N$, we are restricting the difference of the $SU(N)$ and $U(1)$ levels to be an integer multiple of $N$,
\be
k'=k+nN\,,n\in\mathbb{Z}\,.
\ee
This enables us to glue the $U(1)$ and $SU(N)$ gauge fields together to form a gauge invariant theory of a single $U(N)$ gauge field $a=a_{SU(N)}+\tilde{a}\,\mathbf{1}$, with $\Tr[a]=N\tilde{a}$ having quantized fluxes as in Eq. \eqref{eq: flux quantization}. The Lagrangian for the $U(N)_{k,k'}$ theory can be written as
\be
\mathcal{L}_{U(N)_{k,k'}}=\frac{k}{4\pi}\Tr\left[a_{SU(N)}da_{SU(N)}-\frac{2i}{3}a_{SU(N)}^3\right]+\frac{Nk'}{4\pi}\tilde{a}d\tilde{a}\,.
\ee
For the case $k=k'$, we simply refer to the theory as $U(N)_k$. 

Throughout this paper, we implicitly regulate non-Abelian (Abelian) gauge theories using Yang-Mills (Maxwell) terms, as opposed to dimensional regularization \cite{Witten1989,Chen1992}. In Yang-Mills regularization, there is a one-loop exact shift of the $SU(N)$  level, $k\rightarrow k+\operatorname{sgn}(k)N$, that does not appear in dimensional regularization. Consequently, to describe the same theory in dimensional regularization, one must start with a $SU(N)$ level $k_{\mathrm{DR}}=k+\operatorname{sgn}(k)N$. The dualities discussed in this paper, e.g. Eqs.\eqref{eq: U/SU}-\eqref{eq: U/U}, therefore would take a somewhat different form in dimensional regularization.

\end{appendix}

\nocite{apsrev41Control}
\bibliographystyle{apsrev4-1}
\bibliography{pairing}

\end{document}